\documentclass[aps,prl,twocolumn,superscriptaddress,showkeys]{revtex4-1}
\usepackage{graphicx}
\pdfoutput=1

\begin{document}

\preprint{}

\title{Kinetic control of the coverage of oil droplets by DNA-functionalised colloids}


\author{Darshana Joshi}
\affiliation{Cavendish Laboratory, University of Cambridge, Cambridge CB3 0HE,
United Kingdom}
\author{Dylan Bargteil}
\affiliation{Center for Soft Matter Research and Department of Physics, New York University, NewYork, New York, 10003, USA.}
\author{Alessio Caciagli}
\author{Jerome Burelbach}
\author{Zhongyang Xing}
\affiliation{Cavendish Laboratory, University of Cambridge, Cambridge CB3 0HE, United Kingdom}
\author{Andr\'e S. Nunes}
\author{Diogo E. P. Pinto}
\author{Nuno A. M. Ara\'ujo}
\affiliation{Departamento de F\'{\i}sica, Faculdade de Ci\^{e}ncias, and Centro de F\'isica Te\'orica e Computacional, Universidade de Lisboa, Campo Grande, P-1749-016 Lisboa, Portugal.}
\author{Jasna Bruijc}
\affiliation{Center for Soft Matter Research and Department of Physics, New York University, NewYork, New York, 10003, USA.}
\author{Erika Eiser}\email{ee247@cam.ac.uk}
\affiliation{Cavendish Laboratory, University of Cambridge, Cambridge CB3 0HE, United Kingdom}


\date{\today}

\begin{abstract}
We report a study of reversible adsorption of DNA-coated colloids on complementary functionalized oil droplets. We show that it is possible to control the surface coverage of oil droplets by colloidal particles, by exploiting the fact that during slow adsorption, compositional arrest takes place well before structural arrest occurs. As a consequence, we can prepare colloid-coated oil droplets with a `frozen' degree of loading, but with fully ergodic colloidal dynamics on the droplets. We illustrate the equilibrium nature of the adsorbed colloidal phase by exploring the quasi two-dimensional (2d) phase behaviour of the adsorbed colloids under the influence of depletion interactions. We present simulations of a simple model that illustrates the nature of the compositional arrest and the structural ergodicity.
\end{abstract}


\maketitle

\section{Introduction}
One of the key challenges in complex self-assembly is the creation of ordered, quasi-2d patterns of many distinct colloids or nano-particles. To achieve this objective, it is important that the substrate is smooth and clean, and that different species of colloidal particles can bind independently and reversibly to the surface. Moreover, the surface-bound particles should be sufficiently mobile to ensure that the structure that is most stable is also kinetically accessible. 

Most solid substrates are less than ideal for this purpose: the surfaces often contain defects or impurities that trap colloids and, in addition, bound colloids diffuse slowly on such surfaces. Liquid interfaces would be less susceptible to the above problems, however the strength with which colloids bind to liquid-liquid interfaces (through the Pickering mechanism) may be hundred or thousand times stronger than the thermal energy and, as a consequence, controlled and reversible adsorption of different species is difficult to achieve \cite{Ramsden1903, Binks2006, Pickering1907a, Dinsmore2002, Herzig2007}. In addition, control of the interactions between surface-incorporated colloids is difficult because of the dominance of long-ranged capillary forces \cite{Lewandowski2009,Cavallaro2011,Madivala2009}.

Here we introduce a new strategy to assemble colloids reversibly at a liquid-liquid interface. To achieve this we functionalize both the liquid-liquid interface and the colloidal surface, in such a way that colloids can bind reversibly to the interface, through complementary DNA interactions. Of course, DNA has been used extensively as selective glue between nano to micron-sized solid colloids \cite{Mirkin1996, Alivisatos1996,Nykypanchuk2008,Geerts2010}. Other groups have explored functionalizing vesicles or fluid membranes with single-stranded (ss)DNA \cite{Hadorn2012,Parolini2015}. A hybrid approach \cite{VanDerMeulen2013} involved functionalizing fluid membranes enveloping hard spherical colloids, and Feng et al. \cite{Feng2013} have explored the use of DNA-functionalized oil droplets (OD). The motivation for using such fluid substrates was to ensure that the grafted ssDNA could diffuse on the surface. The DNA can then accumulate into `rafts' at the point of contact between two particles with complementary functionalization.  A drawback of the above strategies to achieve mobile grafting was that the amount of ssDNA that could be incorporated into the phospholipid bilayers or monolayers was quite limited \cite{Hadorn2012, Feng2013}. Here we present an approach that overcomes this drawback. This allows us to create oil droplets that are densely grafted with mobile ssDNA. 

In our approach we rely on the creation of oil droplets (ODs) with a typical diameter of $20-30\,\mu m$, stabilised by  sodium dodecylsulfate (SDS), onto which we adsorb polylysine-g[3.5]-polyethyleneglycol-biotin (PLL-PEG-bio), a comb-like polyelectrolyte. As illustrated in Fig. 1 the positively charged poly-lysine backbone adsorbs in a flat manner onto the negatively charged SDS head-groups that are exposed to the oil-water interface.  To every third to fourth lysine repeat unit a $2$ or $3.4\,kD$ PEG chain is attached. Half of the roughly 40 PEG chains per PLL backbone carry a biotin group, which we functionalized with streptavidin from solution. Finally, biotinylated ssDNA, denoted \textbf{A}, was brought onto the OD's surfaces via the streptavidin. Thus we created small flat patches of PLL molecules that diffuse freely with their DNA on the oil-water interface. Adding to the ODs $0.57\,\mu m$ large polystyrene (PS) particles with the complementary ssDNA, \textbf{A'}, these anchor reversibly to the interfaces via DNA hybridization. Depending on the solvent conditions, this seemingly simple system exhibits an unexpectedly rich quasi-2d phase diagram of colloid aggregation.
\section{Results}
\subsection{Preparation of oild droplets (ODs)}
The ODs were prepared in a microfluidic device. Prior to its use the device was plasma cleaned and immediately afterwards filled with water to slow down the loss of hydrophilicity of the device walls, and tested for leaks. The channels used were $20\,\mu m$ wide and $25\,\mu m$ high. Firstly, $5$ to $10\,mM$ SDS solution was flushed in at $250\,\mu L/hr$ to flood the channel. Once SDS has filled the channel, $50\,cSt$ silicone oil ($0.971\,g/mL$) was introduced at $25\,\mu L/hr$ into the channels. Droplets formed at the junction and were collected at the outlet (Fig.1A). We used Nemesys and Harvard pumps. ODs were stored in the fridge in a $10\,mM$ SDS solution -- they remained stable for a minimum of one year. The SDS was purchased from Sigma and used as received. All solutions were prepared in deionized water.
\subsection{DNA Functionalization of ODs}
The coverage of PLL-PEG-bio on the OD was tested by binding fluorescently labelled streptavidin to the biotinylated PEG ends (Fig. 1A(iii)). 
\begin{figure}
\includegraphics[width=0.45\textwidth]{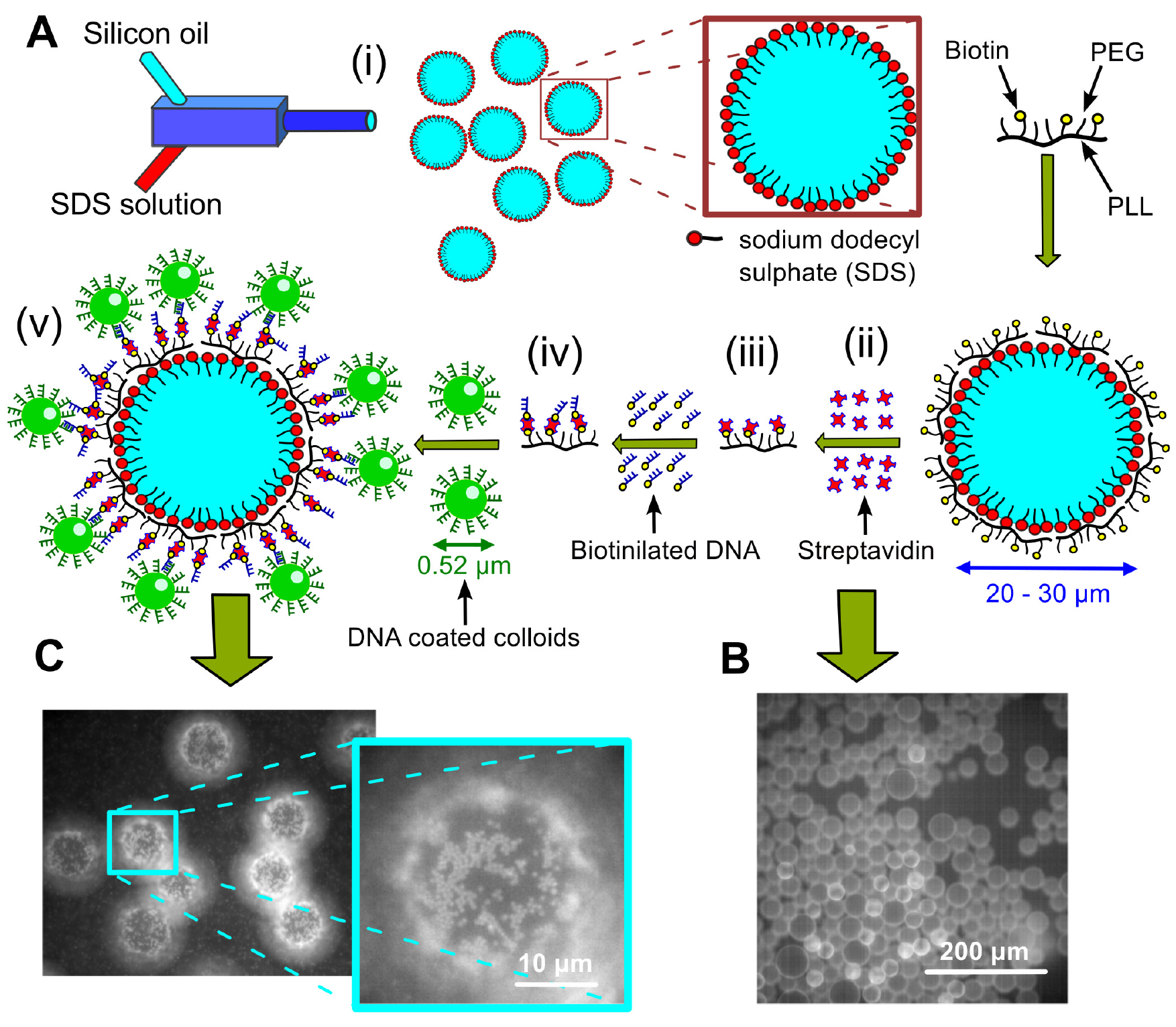}
\caption{A) Cartoon representing various stages of the sample preparation: i) SDS stabilised droplets are prepared by mixing $10\,mM$ SDS and silicone oil in a microfluidic device.(ii) PLL-PEG-Biotin adsorbs flat on the SDS stabilised oil droplets that are negatively charged due to the sulphate head group of the SDS surfactant (iii) Texas Red labelled streptavidin linkers are then attached to the Biotin heads on the ODs from solution, (iv) biotinylated ssDNA (\textbf{A}) is then added attaching to these streptavidin linkers. (v) Green fluorescent polystyrene colloids coated with complementary \textbf{A'} DNA are then allowed to bind by DNA hybridization . (B) Fluorescence images of the oil droplets after attaching the fluorescent streptavidin from solution, and (C) a typical image showing the fluorescent colloids hybridized to the OD surfaces with a zoom onto the \textit{south pole} of the droplet.}
\label{fig:Fig1}
\end{figure}
While a lysine monomer carries one positive charge on the amino group, the repeat-units linked to a PEG chain via that amino group are neutral. Hence, each PLL backbone made of 137 repeat units will carry $\sim$40 positive charges \cite{Rossetti2004}, allowing the backbone to adsorb flat on the negatively charged SDS head-groups at the OD-water interface. The hydrophilic, uncharged and flexible PEG side chains align perpendicular to the interface forming a comb-like geometry \cite{Ruiz-Taylor2001,Rossetti2004}. Note that the biotin-terminated PEG chains are with $3.4\,kDa$ longer than the other ones rendering the biotin accessible. After removing excess PLL-PEG-bio from the OD solution Texas Red labelled streptavidin was added adsorbing readily to the biotinylated PEG end groups. The clearly visible fluorescence on the OD surfaces (Fig. 1B) supports the assumption that the PLL-PEG-bio chains are located at the oil-water interface. We used fluorescence intensity measurements to estimate the maximum PLL-PEG-bio grafting density for a fixed volume of ODs (Fig. S1). All experiments presented here were performed at the onset of the saturated grafting regime to avoid free, unbound chains in solution. An increase in fluorescence intensity by several orders of magnitudes was observed, in comparison to that of biotinylated lipid-monolayer stabilised droplets reported earlier \cite{Feng2013}. Here, we obtained a higher coverage of streptavidin linkers ($\sim10^4\,\mu m^{-2}$) on the OD surface, which corresponds to an average distance of $10\,nm$ between biotin-streptavidin ends. After removing excess streptavidin from the continuous aqueous phase, we grafted biotinylated ssDNA, \textbf{A}, to the OD-solution---again any unbound DNA was removed before adding green fluorescent, $0.52\,\mu m$ large, \textbf{A'} DNA coated polystyrene colloids. Care was taken to keep the final aqueous condition constant with an added NaCl concentration of $\sim50\,mM$. The samples were studied using epifluorescence video microscopy after incubating the samples overnight, allowing for DNA equilibrium-hybridization. In most samples we used colloidal bulk volume fractions of $\Phi_c\leq0.05\%$. Focusing on the 'south pole' of the ODs that are located at the top surface of the sealed capillaries (Fig.1C) it becomes apparent that the PS particles have hybridized to the ODs, while a considerable fraction of the colloids remained in the aqueous bulk phase.

\subsection{Colloidal 2d-aggregation at the oil-water interface}
While keeping the PLL-PEG-bio and \textbf{A} DNA coverage approximately constant for all experiments we studied the effect of varying the SDS concentration on the aggregation behaviour of the PS-colloids hybridized to the ODs (Fig. 2).  
\begin{figure}
\includegraphics[width=0.45\textwidth]{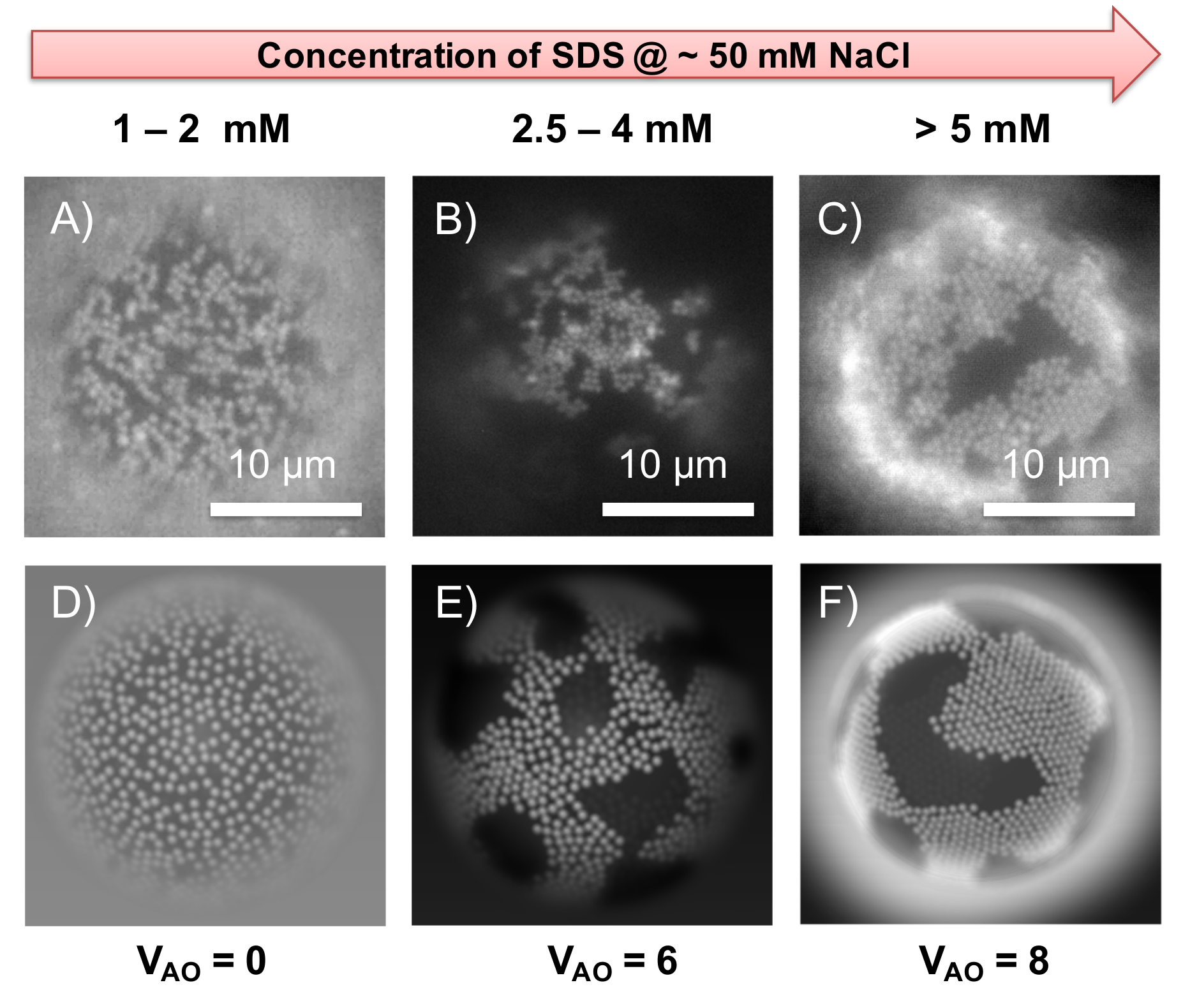}
\caption{Comparison between experimental and numerical results. \textit{Top row}: Fluorescence micrographs (under GFP illumination) showing change in the packing of the $0.52\,\mu m$ polystyrene colloids on the surface of ODs as a function of SDS concentration in the bulk of the sample. The colloids go from A) a colloidal gas-like phase to B) a liquid-like cluster to C) well aligned hexagonal packing at the interface. \textit{Bottom row}: Snapshots for different strengths of the Asakura-Oosawa potential: D)  $V_\mathrm{AO} = 0$, E) 6, and F) 8. These images were modulated with a Gaussian blur emulating the experimental situation.
}
\label{fig:Fig2}
\end{figure}
As we increased the excess SDS concentration in bulk, we observed a transition from a gas-like distribution of hybridized PS-colloids (SDS concentrations $< 2\,mM$; Supporting Video SV 1) to the formation of 2d `continents' with crystalline order at higher concentrations of SDS ($>5\,mM$; SV 2). Between $2\,mM$ and $4\,mM$ SDS concentrations, a two-phase region was observed, where individual hybridized colloids coexisted with more disordered `fluid-like' colloidal islands containing more than just a few colloids (Fig. 2B). We attribute this behavior to depletion attraction induced by SDS micelles. All samples showed a considerable fraction of free, non-hybridized colloids in solution. These did not aggregate because of the steric stabilization provided by the \textbf{A'} DNA brush on the colloids. However, at SDS concentrations above $2\,mM$, we also observed weak colloidal aggregation in bulk, again presumably due to depletion forces.

\subsection{Melting-off colloids from the OD Surfaces}
One of the most striking features of the colloid-OD system is that the loading of oil droplets by colloids appears to depend on the preparation protocol, even though the colloids on the ODs are highly mobile. As we argue below, this protocol dependence is due to the fact that during slow deposition, colloids have more time to `collect' the PLL-PEG-DNA `receptors' than during fast deposition. In fact, during slow deposition, the bound colloids deplete the remaining PLL-PEG-bio domains from the interface, so that no further colloids can bind. 

We first verified whether it was possible to remove the colloids from the OD-water interface by heating the sample well above the hybridization (`melt') temperature, $T_m \simeq 32\,^{\circ}C$, for the sequence used here. The width of the melt region for complementary ssDNA free in solution is typically $\Delta T \sim 10-20\,^{\circ}C$, depending on the solvent conditions and the DNA concentration \cite{DiMichele2014}, but it narrows down to $\sim1\,^{\circ}C$ when the same DNA strands are attached to colloids \cite{Geerts2010}; moreover $T_m$ shifts to higher values. In a control experiment in which we mixed $0.52\,\mu m$ particles functionalized with \textbf{A} or with \textbf{A'} in a 1:1 ratio, using the same solvent condition as those used in the OD experiments, we measured $T_m \simeq 35\,^{\circ}C$ (Fig. S2). 

\begin{figure}
\centerline{\includegraphics[width=0.45\textwidth]{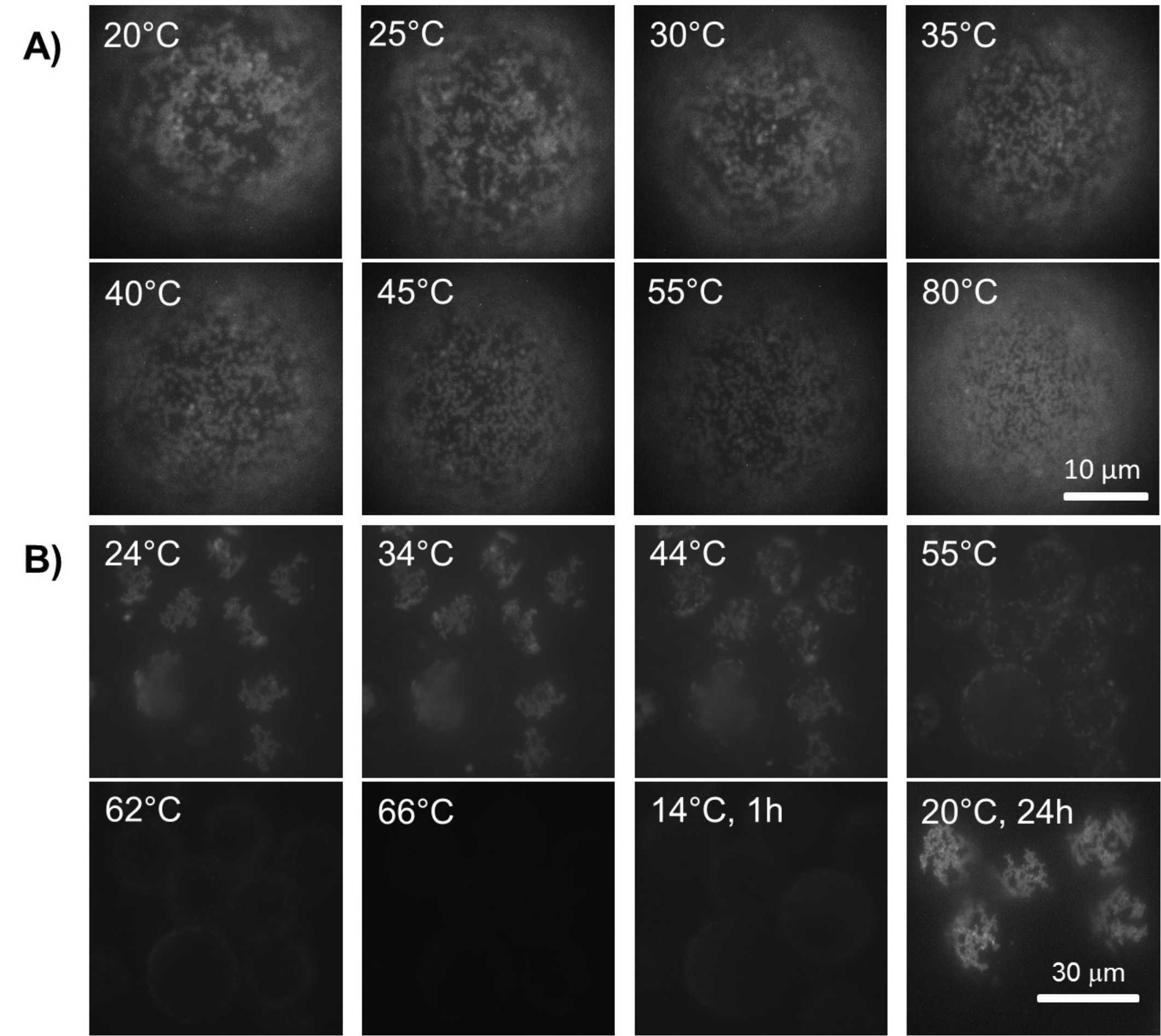}}
\caption{Sequence of fluorescence images heating a sample of $0.57\,\mu m$ colloids grafted to ODs using colloids A) fully covered with \textbf{A'} DNA ($\sim1.5\,mM$) SDS and $50\,mM$ NaCl concentrations); and B) where 3/4 of the \textbf{A'} DNA are replaced by non-binding \textbf{B} ($\sim2.5\,mM$ SDS and $50\,mM$ SDS) . In all images we focus on the \textit{south pole} of the ODs using the same illumination and exposure time. A) For full coverage most colloids remained on the OD surface even when heating to $80\,^\circ C$, though they became increasingly more mobile, which is expressed in the increasingly more blurry appearance of the images. B) For samples with reduced number of binding sites on the colloids melting set in at $\sim 56\,^\circ C$, and at $\sim 70\,^\circ C$ almost all colloids have come off the ODs. The image taken shortly after cooling the sample (over a period of 1 h) to $14\,^\circ C$ showed only few colloids bound to the ODs. The same sample recovered similar coverage and aggregation state as the initial samples after 1 day (bottom most right image).
}
\label{fig:Fig3}
\end{figure}

We first mixed \textbf{A'}-functionalized $0.52\,\mu m$ particles with the \textbf{A}-coated ODs ($\sim 2\,mM$ SDS, $4\,mM$ Tris pH 7-8, $50\,mM$ NaCl) and let the colloids hybridize to the ODs overnight at room temperature. The fluorescence images in Fig. 3A show that at room temperature liquid colloidal islands coexist with non-aggregated colloids hybridized to the interface at this SDS concentration. As we increased the temperature (using a computer-controlled piezoelectric heating stage on the microscope) these islands became smaller and more mobile, but even when heating beyond $35\,^{\circ}C$ the colloids did not detach from the droplet surface --- rather, the particles diffused ever faster. Even at $80\,^{\circ}C$ very few particles came off. We hypothesize that the reason for the strong binding of the colloids is that, as the \textbf{A} DNA on the PLL-PEG is mobile, \textbf{A} DNA will accumulate in the contact region and hence the binding strength will increase with time. Consequently the final number of \textbf{A-A'} bonds in the contact region will be in this case larger (and $T_m$ will be higher) than for the contact region between two hard spheres where the \textbf{A} DNA is fixed. This effect will be enhanced by the fact that the effective contact region between colloids and ODs is larger than between two colloids with the same curvature. However, clearly, the binding strength should still decrease as we decrease the overall \textbf{A'} DNA concentration. To test this hypothesis we reduced the total number of \textbf{A'} strands on the PS particles by replacing 70\% of the \textbf{A'} strands by non-complementary \textbf{B} strands (see Materials \& Methods). The images for increasing temperatures in Fig. 3B show that now particles start to melt off the surface at $\sim56\,^{\circ}C$; at $\sim 66\,^{\circ}C$ almost all bound particles have left the OD surfaces. This supports our assumption that the colloid-OD binding is due to DNA hybridisation and that the very strong binding experienced by colloids with a dense \textbf{A'} coating is due to the accumulation of complementary, PLL-PEG bound DNA in the contact region. We also note that the melting region is not as sharp as for hard particles, although a larger number of binding DNA in the contact area would suggest a narrowing of the melt region. This widening can be understood by considering that the distribution of PLL-PEG bound DNA over the different colloidal contacts is determined by kinetics: some colloids will accumulate more than others. However, once bound, redistribution of PLL-PEG-DNA domains over different colloidal contacts is very slow. Hence, some colloids are more strongly bound than others. This broadening will only occur if the deposition of colloids on the ODs is relatively fast. Otherwise, every colloid that lands on the surface has time to accumulate its full share of PLL-PEG-DNA. But of course, this also means that less PLL-PEG-DNA remains for any subsequent colloids. 

When starting with a high $\Phi_c$, there is no time for the colloids to collect their share of PLL-PEG-DNA and hence more remains for subsequent colloids, and indeed, if we start with a high $\Phi_c$, a higher density of colloids on the ODs is observed, which automatically leads to fewer DNA bridges between any given colloid and the OD, and therefore to a reduced $T_m$. 

Note that after melting off the particles the initial particle grafting density on the OD's is recovered upon cooling to $20\,^{\circ}C$, be it that the process is very slow. An hour after cooling initially to $14\,^{\circ}C$ few colloids have hybridized back to the ODs, and only after 1 day at room temperature we recover the initial colloid density (Fig. 3B).  

\subsection{Simulations}
As it is difficult to probe the distribution of colloid-OD bonds directly in experiments, we used kinetic Monte Carlo simulations to investigate the factors that affect the protocol-dependent OD-colloid binding. In these simulations, the interface is described as a 2d discrete square-lattice with a given density of square patches, $\rho_\mathrm{p}$. A patch represents a single DNA-functionalized PLL-PEG-biotin chain of length $l_\mathrm{p}$, in units of lattice sites, and an interfacial diffusion coefficient $D_\mathrm{p}$. We represent the contact region of the colloids with the interface by a square of length $l_\mathrm{C}$. The number of colloids that arrive per unit time at the oil-water interface is denoted by $F$ (the `flux'). $F$  increases monotonically with the bulk concentration $\Phi_c$. In our simulations we assume that the colloids bind irreversibly if they overlap with at least one site of the patch. Both colloids and patches are assumed to be impenetrable. Since we are interested in the initial stage of binding, we ignore the diffusion of colloid-patch complexes. However, free patches can diffuse and thus accumulate in accessible sites underneath a colloid-patch complex forming a raft. The simulations confirm that as $F$ decreases, the limiting density of surface bound colloids decreases, due to the depletion of accessible patches. 

We also used numerical simulations to study the 2D-aggregation behaviour of the bound colloids on the SDS concentration. To this end, we used Brownian dynamics simulations on a curved interface describing the pair-wise colloid interactions by the superposition of a repulsive Yukawa and the short-ranged, attractive Asakura-Oosawa (AO) potential. Keeping the fraction of bound colloids per OD constant, we increased the strength of the attraction --- it scales linearly with the micelle concentration. Figures 2D-F show snapshots for different strengths of the AO potential. As observed experimentally, by increasing the density of depletant we observe a transition from a fluid-like structure to crystalline-like order (Fig. 2).

\subsection{Particle tracking, differential dynamic microscopy (DDM), microrheology}
To understand how binding to the OD interface affects the dynamics of the colloids compared to free particle motion in bulk we performed particle tracking and DDM measurements. For particle tracking, we used $1.2\,\mu m$ PS colloids coated with a grafting density of \textbf{A'} DNA comparable to the one used on the smaller particles. We also lowered the overall added colloid concentration thus assuring low particle concentrations on the OD surface. We only track trajectories close to the south pole such that the particles move mostly in a plane perpendicular to the viewing direction. We study root mean-square displacements that are much smaller than the radius of the OD and hence we assume that the curvature of the ODs can be neglected in the analysis of the diffusive motion (see Fig. 4A and supporting video SV3; SV4 shows tracking in the bulk). The averaged diffusion coefficient, $D_\mathrm{OD} \sim 4.0\times10^{-14}\,m^2/s^{-1}$, obtained for $1.2\,\mu m$ colloids bound to the ODs was found to be reduced by one order of magnitude compared with that of free particles in solution, $D_\mathrm{sol} \sim 3.37\times10^{-13}\,m^2s^{-1}$; the latter value is close to the theoretical estimate $D_\mathrm{theo} = 4.11\times10^{-13}\,m^2 s^{-1}$ (Fig. 4B). Particle tracking of free PS particles was done in the absence of ODs, but under equivalent solvent conditions and temperature. Feeding the displacements extracted from the same particle tracks into an in-situ analysis program \cite{Yanagishima2011a} we also extracted the effective viscoelastic properties the bound colloids experienced. These suggest that the colloids experience a viscosity of $(9.5 \pm0.6)\,mPa\,s$, which is roughly 10 times larger than that of water. 

\begin{figure}
\centerline{\includegraphics[width=0.45\textwidth]{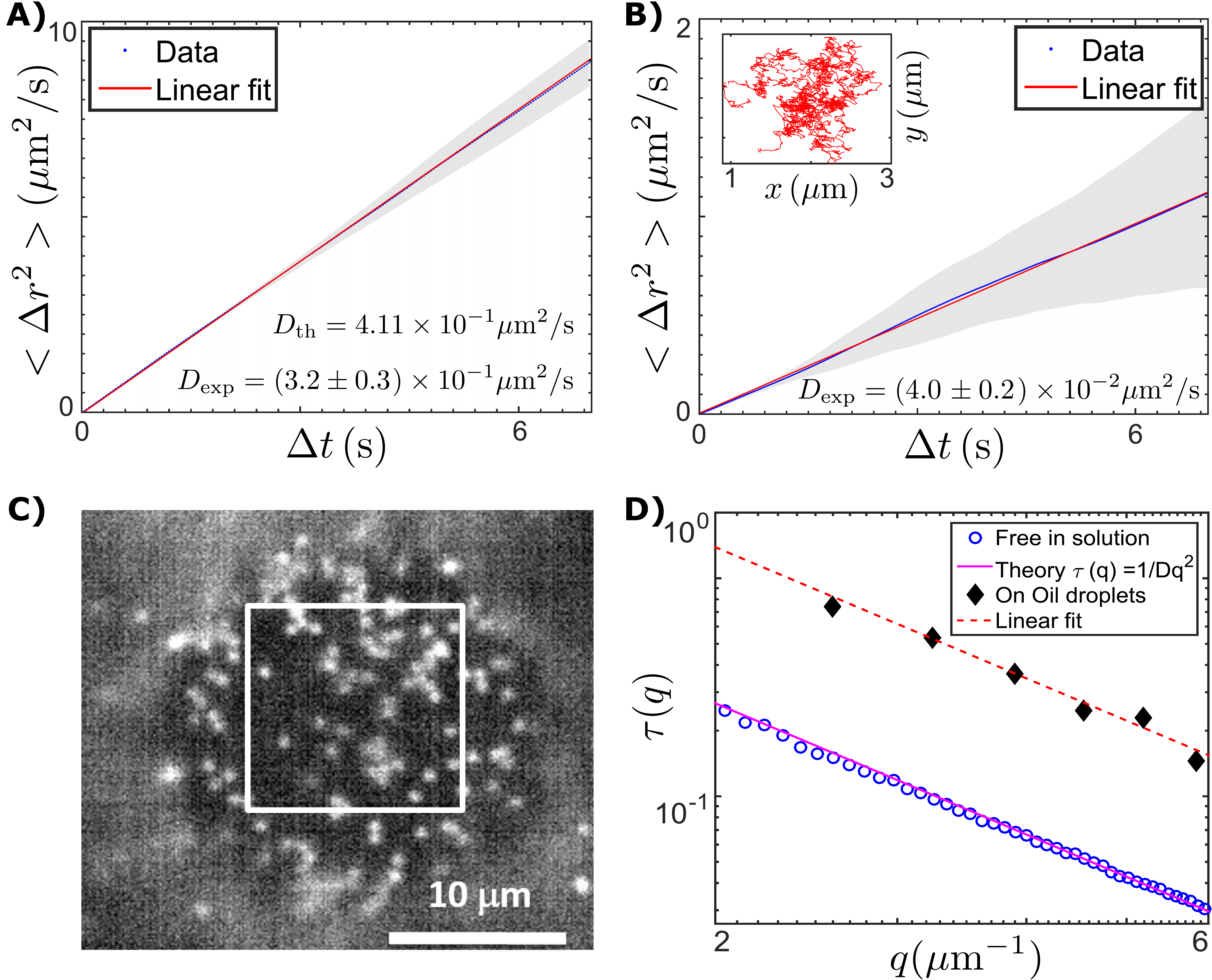}}
\caption{A) Mean square displacement versus delay time  for $1.2\,\mu m$ colloids diffusing free in solution, and B) for DNA-bound colloids diffusing on the OD surfaces averaged over 5 tracks.  The inset shows a typical single-particle track. C) Microscope image of $0.57\,\mu m$ colloids bound to an OD surface focusing on its \textit{south pole}. The square selection was used for DDM analysis. D) Decay time $\tau(q)$ as a function of the scattering vector $q$  extracted from DDM analysis for both colloids diffusing on OD surfaces (as shown in C) and from microscope movies taken from free colloids in solution; the theoretical line passing through the data for the free colloids were obtained assuming a colloid diameter of $0.57\,\mu m$, a viscosity of buffer solution of $0.9\,mPa\,s$, and RT.}
\label{fig:Fig4}
\end{figure}

For samples with only $0.52\,\mu m$ particles present we used Differential Dynamic Microscopy (DDM) to extract the particle diffusivities on the ODs and in bulk \cite{Cerbino2008,Eiser2014}. Similar to the larger particles a considerable reduction in the diffusivity ($\sim2.4\times10^{-13}\,m^2 s^{-1}$) of bound colloids is found, while the value $D_\mathrm{sol} \sim 9.52\times10^{-13}\,m^2 s^{-1}$ measured for the free particles is again similar to the theoretical one (Fig. 4C \& D, and SV1). Note, the DDM results for the particles on the OD surface have a larger error because of the reduced statistics as we limited the sampling to the small region on the south pole to avoid systematic errors due to the curvature of the droplets.

\subsection{Discussion}
The experimental results demonstrate three facts: Firstly that the attachment of our colloids was driven by DNA binding; secondly that this hybridization between colloid and complementary DNA on the surfaces of the ODs was thermally reversible; and thirdly that colloids attached to the surface underwent aggregation, which was purely driven by depletion attraction caused by the surfactant micelles present \cite{Buzzaccaro2007,Buzzaccaro2010,Li2002}. All our experimental findings are supported by the simulation results.

Further proof that our colloids were indeed anchored to the OD-surfaces purely via DNA-hybridization and not due to surface tension effects like in Pickering-Ramsdon emulsions or other non-specific interactions came from several different control experiments. In the first we mixed our OD solutions with colloids coated with non-complementary \textbf{B} DNA, using the standard solvent conditions with an SDS concentration of $\sim2\,mM$. No colloidal adsorption to the ODs was observed, even after a week. In the second control sample we replaced the \textbf{A} DNA on the ODs by PEG-Biotin that formed a similar steric stabilization layer as the DNA linkers, while the colloids were either grafted with \textbf{A'} or \textbf{B} DNA. These samples also did not show any signs of binding between the ODs and the colloids. Even the mixtures of DNA-coated colloids and bare SDS-stabilised ODs did not display any colloidal adsorption on the oil-water interfaces at all bulk SDS-concentrations used. 

In all OD-colloid mixtures, irrespective of the bulk SDS concentration, a large number of colloids remained in solution. As described in the section on melting colloids off the OD-surfaces, we propose that the amount of colloids hybridized to the ODs depends on the total colloid volume fraction $\Phi_c$. Starting with a $\Phi_c \simeq 0.05\,wt\%$, which corresponded roughly to a full DNA coverage of the ODs in solution, we obtained a relatively high colloid coverage (Fig. 1\&2) but still a number of colloids remained free in bulk solution. Using a ten times lower $\Phi_c$, such as used for tracking purposes, we observed much fewer colloids on the OD surfaces. We argue that most of the \textbf{A} DNA bound via the PLL-PEG-bio chains was recruited by the hard colloids forming small rafts of several chains in the contact region. However, when using higher initial colloid concentrations these rafts would be smaller. Once all \textbf{A} DNA was bound no further colloids were able to bind to the oil-water interface. The interfacial regions free of colloids can thus be thought of as regions purely stabilised by SDS. Or in other words, the colloid hybridization leads to segregation into regions of a pure SDS monolayer and others holding the PLL-PEG-bio with the DNA. Because of the fluorescence of the streptavidin attached to the biotins on the PLL-PEG was much weaker than that on the colloids we could not observe this segregation in experiments. However, our simulation studies strongly support our assumptions. 

The strong slowing down of the colloidal diffusion on the ODs also supports our hypothesis that several mobile PLL-PEG-bio chains form rafts. It appears that the measured colloidal diffusion coefficients are dominated by the diffusion coefficient of the rafts. We have shown that the diffusion coefficient of colloids sedimented onto either a `soft' cross-linked polymer surface or a `hard' sterically stabilised support surface does not deviate more than some 10\% from their bulk diffusion value \cite{DiMichele2011}, while here we observed a reduction of one order of magnitude. 

Saffman and Delbr\"{u}ck \cite{Saffman1975} recognized early on that the Brownian motion of proteins attached to a lipid bilayer can be expressed in form of the Stokes-Einstein relation of the translational diffusion coefficient, $D = k_\mathrm{B}Tb$, where $k_\mathrm{B}$ is Boltzmann's constant, $T$ the temperature and $b$ the mobility of the protein (for a free sphere in solution $b = (6\pi\eta R)^{-1}$, with $\eta$ being the viscosity and $R$ the radius of the particles). However, this mobility will be dominated by the viscoelasticity of that layer. They derived an analytical expression for a modified mobility of a cylindrical disk embedded in the double-layer.  Their work was extended to monolayers separating two phases by Fisher et al. and by others, deriving complex expressions and verified in part by simulations \cite{Fischer2006,Fuller2012a}. Sickert and Rondelez reported experimental results on the diffusion of colloidal particles located at the water-air interface separated by lipid monolayers of different densities \cite{Sickert2003}. They found a reduction in the colloids translational diffusion coefficient by one to two orders of magnitude. 

Here we give a rough estimate of the diffusivity of the PLL-PEG-bio rafts. We argue that the PLL chains are in a rather stretched configuration on the surface (Fig. S3) due to steric hindrance between the PEG chains \cite{Rossetti2004,Kawaguchi1997}. The positively charged polylysine will be bound to the SDS surfactants at the water-oil interface, where the hydrophobic tails of the surfactants extend into the roughly 50 times more viscous oil phase. We can estimate the mobility of such a chain by$b = \left(6 \pi \left(\eta_{oil} + \eta_{water} \right) R \right)^{-1}$, where the length of the stretched PLL chains is $\sim48\,nm$, assuming that the length of a lysine monomer is $0.355\,nm$ \cite{Rossetti2004}. This gives us a translational diffusion coefficient of $D_\mathrm{PLL} \simeq 1.8\times10^{-13}\,m^2s^{-1}$, which is about a tenth of that of the free $0.52\mu m$ particles ($D_\mathrm{sol} = 9.52\times10^{-13}\,m^2 s^{-1}$) and surprisingly close to those bound to the surface ($D_\mathrm{OD} \sim 2.4\times10^{-13}\,m^2 s^{-1}$). A slightly larger reduction is observed for the $1.2\mu m$ particles, which is in agreement with our assumption that a larger raft of several PLL-PEG-bio chains may be bound in the contact region. Hence the colloid motion is dominated by the viscous drag of the monolayer-raft with the oil. This finding seems in agreement with the microrheology data, which suggest that the larger colloids experience an apparent viscosity of about $9.\,mPa\,s$, while we extract a viscosity of $\sim 3.5\,mPa\,s$ from the DDM measurements of the bound smaller colloids. 

Our results presented in Fig. 2 demonstrate that the surfactant concentration is an important ingredient in our systems aggregation behaviour. In the final OD-colloid mixture we adjusted the SDS concentration in the aqueous bulk. For the buffer solution with $50\,mM$ added NaCl concentration a minimum of $1-2\,mM$ SDS concentration was required in order to prevent coalescence of the oil droplets. This SDS concentration is just below the Critical Micelle Concentration (CMC) for SDS, which is around $2.5\,mM$ for the ionic strength used in our experiments. Hence, at higher SDS concentrations we create more and more micelles, which induce increasingly stronger depletion attraction between the DNA-stabilised colloids that do not aggregate in the same salty buffer solution but in absence of any surfactant. Depletion attractions between large colloids in bulk solution, induced by small colloids or polymers, have been studied extensively in theory, simulations and experiments \cite{Oosawa1954,Gast1983,Hagen1994,Pusey1993}. Fewer studies used charged surfactant micelles as the depleting agent. Iracki et al. observed the aggregation behavior of `hard', negatively charged silica beads sedimented onto a non-sticking glass surface in the presence of SDS \cite{Iracki2010}. Similar to our experiments they showed that below the CMC hard sphere repulsion dominated, while an increasingly deeper attractive minimum emerged as the SDS concentration increased beyond the CMC. Again, our simulation results confirm that scenario. 

Also, when we go to higher SDS concentrations we see ordered 2d crystals with faceted boundaries. Similar to the observations by Meng et al \cite{Meng2014} made on depletion driven colloidal aggregation on the inside of emulsion droplets these crystals remain finite size and `fracture' due to the competition between elastic deformation of the 2d crystals that prefer a flat arrangement and the binding strength of the colloids to the interface via DNA. As the melting experiments demonstrate any clusters formed at room temperature will become smaller and more mobile upon heating. To summarize, we have introduced a new colloid aggregation mechanism on emulsion droplets that allow reversible detachment from the OD surface in a controlled manner using DNA hybridization. The fact that the hybridization recruits the mobile linkers between the colloids and the OD may be used as model system to understand and study the dynamics of particle or protein adsorption to biological molecules in a crowded environment. The very slow dynamics of the aggregation process may help understand how this aggregation can be controlled \cite{Marenduzzo2006,Ruiz-Taylor2001}.

\section{Materials and Methods}
\subsection{\textit{Coating droplets and colloids with DNA}}
{\small
The DNA strands used in experiments have been purchased from Integrated DNA Technology (IDT). The three sticky ends are \textbf{A} = 5'--CCG GCC--3', \textbf{A'} = 5'--GGC CGG--3', \textbf{B} = 5'--CG CAG CAC C--3'. The \textbf{XX'} pair acted as rigid double-stranded spacer as described in earlier work \cite{Varrato2012}. They were hybridized prior to being attached to either the oil droplets (\textbf{A}) or the colloids (\textbf{A'}). 

In order to achieve a similar DNA-binding capacity/$\mu m^2$ on the oil droplet as that on the colloids, $50\,\mu l$ of the freshly prepared oil droplets stabilised in $10\,mM$ SDS solutions were taken from the creamed layer and mixed with $22.5\,\mu l$ ($1\,mg/mL$) of PLL (20)-g[3.5]- PEG(2)/PEG(3.4)- biotin(50\%)  (purchased from SuSoS AG, Switzerland) and $250\,\mu l$ of $1\,mM$ SDS, $4\,mM$ Tris buffer pH 7.5, maintaining an overall $50\,mM$ concentration of NaCl and $>2\,mM$ concentration of SDS. This mixture was incubated on rollers overnight and washed twice with a wash solution made of $5\,mM$ SDS, $50\,mM$ NaCl, $4\,mM$ Tris pH 7-8, and then dispersed in a suspending solution ($2\,mM$ SDS, $4\,mM$ Tris pH 7-8, $50\,mM$ NaCl). Since the droplets float, we used a syringe to remove the buffer beneath the droplets. Then, $3\,\mu l$ ($1\,mg/ml$) of Texas Red labelled streptavidin (Sigma-Aldrich) per $50\,\mu l$ of the droplet solution was added and incubated on rollers for another hour. Again the droplets were washed twice with the wash solution and suspended in suspending solution. This procedure provided oil droplets with approximately the same streptavidin coverage ($\sim10^4 /\mu m^2$) as that on the colloids. 

The hybridized DNA constructs with double stranded spacers were grafted onto the oil droplets and the colloids using the biotin-streptavidin binding \cite{DiMichele2013}. \textbf{A} DNA was added to the suspension of streptavidin-coated oil droplets in 10x excess. They were allowed to bind on rollers at a $50\, mM$ NaCl and $2\,mM$ SDS solution.  After 4-6 hours the salt concentration was raised to $100\,mM$, while maintaining the SDS concentration. After another 3 hours on the rollers the eppendorf tubes were allowed to stand vertically so that the oil droplets could float to the top and excess DNA and salt could be removed. The DNA coated ODs were then washed thrice using the wash solution and resuspended in the suspending solution. 

$5\,\mu l$ of Green fluorescent colloids ($D\sim0.5\,\mu m$, from Microparticles GmbH) were diluted to $400\,\mu l$ by adding TE buffer and then sonicated for 30 minutes before adding \textbf{A'} DNA. $1\,M$ concentrate of the NaCl solution was then added to this mixture establishing an overall salt concentration of $100\,mM$. After incubating on the rollers for 4-6 hours more NaCl solution was added raising the concentration to $300\,mM$ of NaCl. This was allowed to stay on the rollers for another 4 hours. The colloids were then pelleted using a micro centrifuge, the supernatant was removed and the DNA coated colloids were then resuspended in fresh TE buffer (heated to $40\,^{\circ}C$). This washing protocol was repeated 4-6 times thus removing excess DNA and salt present in the system \cite{DiMichele2013,DiMichele2014}. The last two washes were done in TE containing $100\,mM$ NaCl thereby ensuring steric stability of the DNA brush attached to the colloids. 

Finally OD and colloid solutions were mixed and allowed to bind overnight on rollers --- the final NaCl concentration used in all experiments was $50\,mM$. The desired SDS concentration was adjusted at this stage. }
\subsection{\textit{Sample chambers}}
{\small
Capillary chambers purchased from CM scientific ($0.2\times4\,mm$ ID) were irradiated under a UV lamp for 30 minutes and then plasma cleaned in an oxygen plasma oven for 2 minutes (Diener Elecronics Femto). About $40\,\mu l$ of the sample was filled into the capillary chambers using pipettes and then sealed and glued onto a glass slide using a two-component epoxy glue.}

\subsection{\textit{Imaging and temperature cycling}}
{\small
For imaging we used a Nikon Ti-E inverted epifluorescence microscope with either a Nikon Plan Fluor E40x/0.75 dry objective, or Nikon Plan APO 60x/1.20 water immersion objective, and a Grasshopper3, Point Grey Research Inc., Sony IMX174 CMOS sensor CCD camera and a `perfect focus system' allowing for long tracking measurements. A blue LED source (Cree XPEBLU, $485\,nm$) and white light in combination with a filter cube were used to excite the fluorescence on the colloids and ODs. The sample temperature was controlled via a computer using a home-made Peltier-stage with a thermocouple. Typical heating and cooing rates were $1\,^{\circ}C/min$.}

\subsection{\textit{Image analysis and diffusivity measurements}}
{\small
Image conversion and analysis were done using customised ImageJ and Matlab routines. Single particle diffusivity movies were analysed and tracks were generated using the ANALYSE subroutines of ImageJ. These tracks were subsequently analysed in Matlab generating mean-squared displacements to calculate the diffusivities. Several bright-field time series of 1-2 minute durations were recorded and evaluated using a Matlab routine for DDM analysis developed by S. H. Nathan \cite{Eiser2014,Nathan2015}. In DDM, bright field (or fluorescence) microscope images separated by a given time lag are subtracted such that only the dynamic information due to colloid motion remains. Fourier transforming these difference images for varying time lags and correlating them provides the systems relaxation time, $\tau = (Dq^2)^{-1}$, as function of the scattering wavelength $q$.}

\section{Acknowledgments}
AC acknowledges the ETN-COLLDENSE (H2020-MCSA- ITN-2014, Grant No. 642774). EE and JB (Burelbach) thank the Winton Program for the Physics of Sustainability for the Pump Prime Grant and the scholarship award respectively. DJ thanks the Udayan Care - Vcare grant, the Nehru Trust for Cambridge University, Schlumberger Foundation's FFTF program and Hughes Hall - Santander Bursary. ZX thanks the NUDT Scholarship at Cambridge. EE thanks M. Muthukumar for discussions. AN, DP, and NA acknowledge discussions with M. Telo da Gama, and financial support from the Portuguese Foundation for Science and Technology (FCT) - EXCL/FIS-NAN/0083/2012, UID/FIS/00618/2013, and IF/00255/2013. JB (Bruijc) thanks the Materials Research Science and Engineering Center (MRSEC) program of the National Science Foundation under Award Number DMR-1420073.
\bibliography{basename of .bib file}

\begin{thebibliography}{10}

\bibitem{Ramsden1903}
Ramsden W
\newblock (1903) {Separation of Solids in the Surface-Layers of Solutions and
  'Suspensions' (Observations on Surface-Membranes, Bubbles, Emulsions, and
  Mechanical Coagulation) - Preliminary Account}.
\newblock \emph{Proc. R. Soc. London} 72:156--164.

\bibitem{Binks2006}
Binks BP, Murakami R
\newblock (2006) {Phase inversion of particle-stabilised materials from foams
  to dry water}.
\newblock \emph{Nat. Mater.} 5:865--869.

\bibitem{Pickering1907a}
Pickering SU
\newblock (1907) {CXCVI Emulsions}.
\newblock \emph{J. Chem. Soc. Trans.} 91:2001.

\bibitem{Dinsmore2002}
Dinsmore AD, {et~al.}
\newblock (2002) {Colloidosomes: selectively permeable capsules composed of
  colloidal particles.}
\newblock \emph{Science} 298:1006--9.

\bibitem{Herzig2007}
Herzig EM, White KA, Schofield AB, Poon WCK, Clegg PS
\newblock (2007) {Bicontinuous emulsions stabilised solely by colloidal
  particles.}
\newblock \emph{Nat. Mater.} 6:966--971.

\bibitem{Lewandowski2009}
Lewandowski EP, Bernate JA, Tseng A, Searson PC, Stebe KJ
\newblock (2009) {Oriented assembly of anisotropic particles by capillary
  interactions}.
\newblock \emph{Soft Matter} 5:886.

\bibitem{Cavallaro2011}
Cavallaro M, Botto L, Lewandowski EP, Wang M, Stebe KJ
\newblock (2011) {From the Cover: Curvature-driven capillary migration and
  assembly of rod-like particles}.
\newblock \emph{Proc. Natl. Acad. Sci.} 108:20923--20928.

\bibitem{Madivala2009}
Madivala B, Vandebril S, Fransaer J, Vermant J
\newblock (2009) {Exploiting particle shape in solid stabilised emulsions}.
\newblock \emph{Soft Matter} 5:1717.

\bibitem{Mirkin1996}
Mirkin C, Letsinger R, Mucic R, Storhoff J
\newblock (1996) {A DNA-based method for rationally assembling nanoparticles
  into macroscopic materials}.
\newblock \emph{Nature} 382:607--609.

\bibitem{Alivisatos1996}
Alivisatos A, Johnsson K, Peng X
\newblock (1996) {Organization of'nanocrystal molecules' using DNA}.
\newblock \emph{Nature} 382.

\bibitem{Nykypanchuk2008}
Nykypanchuk D, Maye MM, {Van der Lelie} D, Gang O
\newblock (2008) {DNA-guided crystallization of colloidal nanoparticles}.
\newblock \emph{Nature} 451:549--552.

\bibitem{Geerts2010}
Geerts N, Eiser E
\newblock (2010) {DNA-functionalized colloids: Physical properties and
  applications}.
\newblock \emph{Soft Matter} 6:4647.

\bibitem{Hadorn2012}
Hadorn M, {et~al.}
\newblock (2012) {Specific and reversible DNA-directed self-assembly of
  oil-in-water emulsion droplets}.
\newblock \emph{Proc. Natl. Acad. Sci.} 109:20320--20325.

\bibitem{Parolini2015}
Parolini L, {et~al.}
\newblock (2015) {Volume and porosity thermal regulation in lipid mesophases by
  coupling mobile ligands to soft membranes}.
\newblock \emph{Nat. Commun.} 6:5948.

\bibitem{VanDerMeulen2013}
Van~der Meulen SAJ, Leunissen ME
\newblock (2013) {Solid colloids with surface-mobile DNA linkers}.
\newblock \emph{J. Am. Chem. Soc.} 135:15129--15134.

\bibitem{Feng2013}
Feng L, Pontani L, Dreyfus R, Chaikin P, Brujic J
\newblock (2013) {Specificity, flexibility and valence of DNA bonds guide
  emulsion architecture}.
\newblock \emph{Soft Matter} 9:9816.

\bibitem{Rossetti2004}
Rossetti FF, {et~al.}
\newblock (2004) {Interaction of Poly(L-Lysine)-g-Poly(Ethylene Glycol) with
  Supported Phospholipid Bilayers}.
\newblock \emph{Biophys. J.} 87:1711--1721.

\bibitem{Ruiz-Taylor2001}
Ruiz-Taylor LA, {et~al.}
\newblock (2001) {Monolayers of derivatized poly(L-lysine)-grafted
  poly(ethylene glycol) on metal oxides as a class of biomolecular interfaces}.
\newblock \emph{Proc. Natl. Acad. Sci.} 98:852--857.

\bibitem{DiMichele2014}
Michele LD, {et~al.}
\newblock (2014) {Effect of Inert Tails on the Thermodynamics of DNA
  Hybridization}.
\newblock \emph{J. Am. Chem. Soc.} 136:6538--6541.

\bibitem{Yanagishima2011a}
Yanagishima T, Frenkel D, Kotar J, Eiser E
\newblock (2011) {Real-time monitoring of complex moduli from micro-rheology}.
\newblock \emph{J. Phys. Condens. Matter} 23:194118.

\bibitem{Cerbino2008}
Cerbino R, Trappe V
\newblock (2008) {Differential Dynamic Microscopy: Probing Wave Vector
  Dependent Dynamics with a Microscope}.
\newblock \emph{Phys. Rev. Lett.} 100:188102.

\bibitem{Eiser2014}
Eiser E
\newblock (2014) in \emph{Multi Length-Scale Characterisation}, eds{} {Duncan
  W. Bruce} DO, Walton RI
\newblock (John Wiley {\&} Sons, Ltd.) No.{} May, pp 233--282.

\bibitem{Buzzaccaro2007}
Buzzaccaro S, Rusconi R, Piazza R
\newblock (2007) {"Sticky" Hard Spheres: Equation of State, Phase Diagram,
  and Metastable Gels}.
\newblock \emph{Phys. Rev. Lett.} 99:098301.

\bibitem{Buzzaccaro2010}
Buzzaccaro S, Colombo J, Parola A, Piazza R
\newblock (2010) {Critical Depletion}.
\newblock \emph{Phys. Rev. Lett.} 105:198301.

\bibitem{Li2002}
Li W, Ma HR
\newblock (2002) {Depletion potential near curved surfaces}.
\newblock \emph{Phys. Rev. E} 66:061407.

\bibitem{DiMichele2011}
Michele LD, {et~al.}
\newblock (2011) {Interactions between colloids induced by a soft cross-linked
  polymer substrate}.
\newblock \emph{Phys. Rev. Lett.} 107:1--4.

\bibitem{Saffman1975}
Saffman PG, Delbruck M
\newblock (1975) {Brownian motion in biological membranes}.
\newblock \emph{Proc Natl Acad Sci USA} 72:3111--3113.

\bibitem{Fischer2006}
Fischer TM, Dhar P, Heinig P
\newblock (2006) {The viscous drag of spheres and filaments moving in membranes
  or monolayers}.
\newblock \emph{J. Fluid Mech.} 558:451.

\bibitem{Fuller2012a}
Fuller GG, Vermant J
\newblock (2012) {Complex Fluid-Fluid Interfaces: Rheology and Structure}.
\newblock \emph{Annu. Rev. Chem. Biomol. Eng.} 3:519--543.

\bibitem{Sickert2003}
Sickert M, Rondelez F
\newblock (2003) {Shear viscosity of langmuir monolayers in the low-density
  limit.}
\newblock \emph{Phys. Rev. Lett.} 90:126104.

\bibitem{Kawaguchi1997}
Kawaguchi S, {et~al.}
\newblock (1997) {Aqueous solution properties of oligo- and poly(ethylene
  oxide) by static light scattering and intrinsic viscosity}.
\newblock \emph{Polymer (Guildf).} 38:2885--2891.

\bibitem{Oosawa1954}
Oosawa F, Asakura S
\newblock (1954) {Surface Tension of High-Polymer Solutions}.
\newblock \emph{J. Chem. Phys.} 22:1255.

\bibitem{Gast1983}
Gast AP, Hall CK, Russel WB
\newblock (1983) {Polymer-induced phase separations in nonaqueous colloidal
  suspensions}.
\newblock \emph{J. Colloid Interface Sci.} 96:251--267.

\bibitem{Hagen1994}
Hagen MHJ, Frenkel D
\newblock (1994) {Determination of phase diagrams for the hard-core attractive
  Yukawa system}.
\newblock \emph{J. Chem. Phys.} 101:4093.

\bibitem{Pusey1993}
Pusey PN, Pirie AD, Poon WCK
\newblock (1993) {Dynamics of colloid-polymer mixtures}.
\newblock \emph{Phys. A Stat. Mech. its Appl.} 201:322--331.

\bibitem{Iracki2010}
Iracki TD, Beltran-Villegas DJ, Eichmann SL, Bevan MA
\newblock (2010) {Charged Micelle Depletion Attraction and Interfacial
  Colloidal Phase Behavior}.
\newblock \emph{Langmuir} 26:18710--18717.

\bibitem{Meng2014}
Meng G, Paulose J, Nelson DR, Manoharan VN
\newblock (2014) {Elastic Instability of a Crystal Growing on a Curved
  Surface}.
\newblock \emph{Science (80-. ).} 343:634--637.

\bibitem{Marenduzzo2006}
Marenduzzo D, Finan K, Cook PR
\newblock (2006) {The depletion attraction: an underappreciated force driving
  cellular organization}.
\newblock \emph{J. Cell Biol.} 175:681--686.

\bibitem{Varrato2012}
Varrato F, {et~al.}
\newblock (2012) {Arrested demixing opens route to bigels}.
\newblock \emph{Proc. Natl. Acad. Sci.} 109:19155--19160.

\bibitem{DiMichele2013}
Michele LD, {et~al.}
\newblock (2013) {Multistep kinetic self-assembly of DNA-coated colloids}.
\newblock \emph{Nat. Commun.} 4:2007.

\bibitem{Nathan2015}
Nathan S
\newblock (2015) Ph.D. thesis (University of Cambridge).

\bibitem{Ramos1997}
Ramos L, Fabre P
\newblock (1997) {Swelling of a Lyotropic Hexagonal Phase by Monitoring the
  Radius of the Cylinders}.
\newblock \emph{Langmuir} 13:682--686.

\end{thebibliography}

\end{document}